\documentclass[twocolumn,preprintnumbers,amsmath,amssymb]{revtex4}
\usepackage{epsfig}
\usepackage{amsmath}
\graphicspath{{figures/}}
\usepackage{epsfig}
\usepackage{amsmath}
\usepackage{color}
\usepackage{graphicx,epsfig}
\usepackage{epstopdf}
\usepackage{breqn}
\usepackage{multirow}

\AppendGraphicsExtensions{.tif}
\usepackage{color}
\usepackage{float}
\usepackage{natbib}
\usepackage{hyperref}
\usepackage{subfigure}

\begin{document}

\title{Light-to-heat conversion of optically trapped hot Brownian particles.}

\author{Elisa Ortiz-Rivero$^{1,\dag}$, Sergio Orozco-Barrera$^{2,\dag}$, Hirak Chatterjee$^{2}$, Carlos D. Gonz\'alez-G\'omez$^{2,3}$, Carlos Caro$^{4,5}$, Mar\'ia L. Garc\'ia-Mart\'in$^{4,5,6}$, Patricia Haro-Gonz\'alez$^1$, Ra\'ul A. Rica$^{2,*}$, Francisco G\'amez$^{7,*}$.\\
\vspace{0.5 cm} $^1$ Nanomaterials for Bioimaging Group,  Departamento de F\'isica de Materiales, Facultad de Ciencias \& Instituto Nicol\'as Cabrera Universidad Aut\'onoma de Madrid, Madrid 28049, Spain.\\
$^2$ Universidad de Granada, Nanoparticles Trapping Laboratory, Research Unit Modeling Nature (MNat) and Department of Applied Physics, 18071 Granada, Spain.\\
$^3$ Universidad de M\'alaga, Department of Applied Physics II, 29071 M\'alaga, Spain.\\
$^4$ Biomedical Magnetic Resonance Laboratory-BMRL, Andalusian Public Foundation Progress and Health-FPS, Sevilla, Spain.\\
$^5$ Biomedical Research Institute of M\'alaga and Nanomedicine Platform (IBIMA-BIONAND Platform), University of M\'alaga, C/Severo Ochoa 35, 29590 M\'alaga, Spain.\\
$^6$ Biomedical Research Networking Center in Bioengineering, Biomaterials \& Nanomedicine (CIBER-BBN),
28029 Madrid, Spain.\\
$^{7}$ Department of Physical Chemistry, Universidad Complutense de Madrid, 28040 Madrid, Spain.\\
*Corresponding authors: rul@ugr.es, frgamez@ucm.es}

\begin{abstract} 

Anisotropic hybrid nanostructures stand out as promising therapeutic agents in photothermal conversion-based treatments. Accordingly, understanding local heat generation mediated by light-to-heat conversion of absorbing multicomponent nanoparticles at the single particle level have both forthwith become a subject of broad and current interest. Nonetheless, evaluating reliable temperature profiles around a single trapped nanoparticle is challenging from all experimental, computational, and fundamental viewpoints. Committed to filling this gap, the heat generation of an anisotropic hybrid nanostructure is explored by means of two different experimental approaches from which the local temperature is measured in a direct or indirect way, all in the context of the hot Brownian motion theory. The results were compared with analytical results supported by the numerical computation of the wavelength-dependent absorption efficiencies in the discrete dipole approximation for scattering calculations, which has been here extended to inhomogeneous nanostructures. Overall, we provide a consistent and comprehensive view of the heat generation in optical traps of highly absorbing particles under the viewpoint of the hot Brownian motion theory. \\

\end{abstract}

\maketitle

\noindent \textbf{\textit{Keywords}}. Optical tweezers, hybrid nanostructures, heat generation, nanothermometry, hot Brownian motion.\\

\def\thefootnote{\dag}\footnotetext{These authors contributed equally to this work}\def\thefootnote{\arabic{footnote}}

\section{Introduction}

Multicomponent nanoparticles (MCNPs) can be defined as single structures that combine the properties of at least two different materials at the nanoscale \cite{Caro21,Ximendes21,Ha19}. A myriad of applications of à la carte MCNPs have occupied the limelight of different research fields throughout the last years because of the virtually endless materials combination that confers specific features designed for each potential purpose \cite{Diez22}. For instance, their relevance in biomedical applications has taken off due to their ability to act both as a dual agent for multimodal contrast and/or for combined/synergic therapies \cite{Yankeelov14,Bayat17}. In this context, materials engineering has developed specific routes for costuming a wide plethora of on-demand nanocomposites whose shape, functionalization and spectroscopic attributes of the surface plasmon resonance (SPR) provide distinctive physical and biocompatibility features that are key for real-world applications in nanomedicine \cite{Nguyen18,Ding17}. The so-called inductive synthesis has emerged in this field as a new and intriguing branch of nanoscience that foster the adaptation of the morphology of nanoparticles to achieve specific biological responses \cite{Zhang22}. In particular, MCNPs comprised by a light-absorbing material (\textit{i.e.}, gold) and magnetic moieties can be conceived as agents for hyperthermia treatments by dint of their efficient heat-conversion effect under both alternating magnetic fields and near-infrared light \cite{baffou2013thermo,Tancredi19,Zhang16,Tarkistani21}. It becomes then apparent that measuring temperatures at the nanoscale and understanding the photothermal conversion of light-absorbing hybrid nanoparticles at the single particle level have become a central issue in applied physics and materials engineering.\\

A captivating approach to manipulate, control and measure properties of isolated nanostructures is embodied by the continuously upgraded technology of optical tweezers \cite{gieseler2021optical}. Even if challenging, trapping absorbing nanoparticles, in particular those made of noble metals that feature plasmonic resonances, has been recently optimized \cite{oro,kang12,review,odebo2020optical}. Particularly, optical tweezers paved the way towards the microscopic comprehension of heat generation by isolated nanoparticles by monitoring changes in the properties of the heating source \cite{Andres-Arroyo15,Andren17} or its surroundings \cite{siler16}. Some previous works were devoted to evaluating the local temperature in optically trapped nanoparticles by means of viscosity variations of the surrounding media upon laser heating \cite{Paloma2018,MikaelNanorod2017,jungova} or by the indirect monitorization of the temperature-dependent emission of a nanoprobe \cite{Paloma2016}. In the last case, the eventual influence of the neighbouring nanothermometers on the absorption, the center of mass motions and the light-to-heat conversion efficiency of the trapped particle is a question that should be considered. But a word of caution, temperature is a many-particle property, and therefore it is a blurry-defined concept from statistical mechanics grounds in single-particle systems far from equilibrium, jeopardising the validity of traditional Brownian motion prescriptions. In consequence, another point of interest is whether the measured "temperature" resembles the classic interpretation of surface or internal temperatures or if it is just a value that averages local thermal fluctuations of the solvent bath within the trap volume enabling the description of the particle dynamics within the so-called hot Brownian motion (HBM) theory \cite{rings10,millen14,riviere2022hot,HBM3}. To address these points, experimental and theoretical results on the heat generation and local temperature of a light-absorbing multicomponent nanoparticle under optical trapping conditions are accomplished here.\\

\section{Results and discussion}

\subsection{Characterization of the MCNPs}

\subsubsection{Physicochemical characterization}

In this work, we faced thermometric measurements on hybrid nanoflowers (NFs) synthesized by the multinucleation of magnetite globular petals on spherical gold seeds, as reported in reference \cite{Christou22} and detailed in the Experimental Section. Transmission electronic microscopy (TEM) was employed to characterize the MCNPs morphology and the size distribution of the moieties of each component. The characteristic shapes are presented in the micrography of Figure \ref{fig1}(a). The diameter of the different inorganic counterparts were $\sim$12 nm for both the Au core and Fe$_3$O$_4$ petals, respectively, as shown in the size histograms shown in Figure S1. Overall, and assuming \textit{effective} spherical shape, the TEM diameter of the whole nanoflower was $\sim$33 nm. After a PEGylation procedure, the colloidal stability of the aqueous suspension was confirmed by both the constancy of the hydrodynamic diameter ($\sim$58 nm) for one month and the negative value of the nanoparticle $\zeta$-potential of --14 mV that, according to DLVO theory, was expected to lead to stable suspensions \cite{Campbell}. The extinction spectra of a diluted suspension, shown in Figure \ref{fig1}(b), presented a broad band with a maximum extinction wavelength at $\sim$535 nm on account of the SPR of the gold core perturbed by the interaction with the magnetite globules. The light extinction increases in the NIR from $\sim$800 nm on. This behavior has been reported before in both experiments and Mie calculations and is ascribed to a charge-transfer absorption band in the magnetic moiety \cite{magnetiteNIR}. The behavior of the long-tail extinction of the particles in the NIR envelops both the first and second biological windows \cite{Hemmer16} that, together with their watery stability, enable their fundamental optical exploration as hyperthermia agents \cite{Li20}. From a purely optical vista, these nanostructures combine the penetration depth and absorption of magnetite with the reflecting, heating and trapping advantages of gold. \\


\subsubsection{Evaluation of the absorption cross section from DDA simulations}

\begin{figure*}[!ht]\centering\hspace*{0cm} \includegraphics[width=18cm, angle= 0]{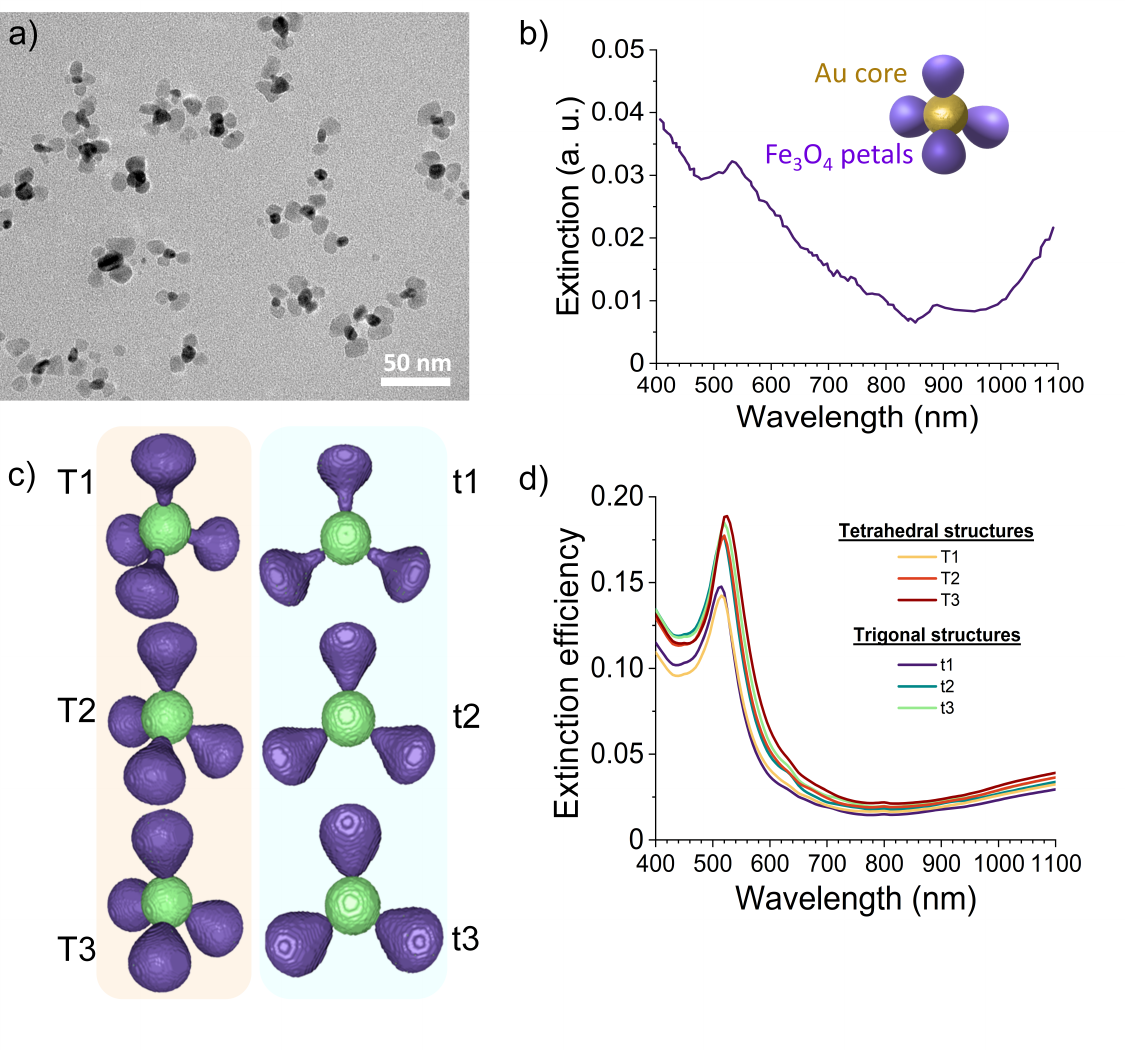}
\vspace*{-0.5cm} \caption{(a) TEM image of the synthesized nanoflowers. (b) Extinction spectra of a diluted suspension of the nanoflower suspensions. The inset represents an schematic model of the composition of the nanoflowers. (c) Theoretical models for the calculation of the scattering profiles of the nanoflowers: models of the whole nanoflowers with 4 (left column) and 3 (right column) petals, for $\beta=1.5$ (T1, t1), $\beta=2.5$ (T2, t2), and $\beta=5$ (T3, t3). (d) Extinction efficiencies for all the structures calculated with DDA.} 
\label{fig1}
\end{figure*}

The synthesized nanoflowers were simulated using parametric CAD modelling. To define the nanoflower geometry, the petals have been modelled using a piriform curve, which is parameterized just by one single parameter $\beta$ (see Figure S2 in the Supplementary Material). Sweeping the value of $\beta$ from 1.5 to 6 with 0.5 step size, we developed the parametric curves of the profiles of the petals, which were revolved in three dimensions to obtain the shapes of the petals, as shown in Figure S2. The flower shape has been conceived by the assignment of the petals around the core in both planar triangular and tetrahedral arrays for three selected shapes of petals having $\beta$ values 1.5, 2.5 and 5. Since the rise in $\beta$ leads to a concomitant increase in the height of the petals, a scaling factor has been introduced to keep the size ratio of the core and petal within the limits of the experimental size range as shown by the histogram in Figure S1. The diameters of the core gold and the magnetite caps are set to 10 nm and 11.5 nm, respectively. Corresponding models for the triangular and tetrahedral arrays are shown in Figure \ref{fig1}(c). Models (T1) and (t1) are the tetrahedral and triangular nanoflower for $\beta=1.5$. Similarly, models (T2) and (t2) show the tetrahedral and triangular nanoflower for $\beta=2.5$, and models (T3) and (t3) show the tetrahedral and triangular nanoflower for $\beta=5$, respectively. The compositional assignment of the models for the particles synthesized in the present work has been modelled using a home-built Python code \cite{hirak_git}. In view of the symmetric structure of the bimetallic nanoflower, the design of the code involves superposition of magnetite petals with the gold core seeded inside. The interface between the two dielectrics has been set to the dipole distance between the dipolar lattice considered for the scattering simulation considering discrete dipole approximation (DDA) for gold core and magnetite petals. The scattering simulation has been performed using DDSCAT code to calculate scattering, absorption and extinction cross-section at different wavelengths with corresponding nearfield calculations within $\pm$2 nm from the gold core in both $x$- and $y$- axis  \cite{DDSCAT}. The result illustrates a strong electromagnetic field appearing at the interface between the gold core and magnetite petals, as shown in Figure S3 for model (T3). The appearance of strong electric field indicates strong confinement of hot electrons at the dielectric boundary, leading to processes like hot electron transfer or plasmon-phonon coupling, among others, which could open up a future prospect of study. Finally, a plot of DDA calculated extinction efficiencies for the six models is shown in Figure~\ref{fig1}(d). The patterns show a sharp peak at around 500 nm and a shoulder tail around 1100 nm, depicting the inclusion of both gold and magnetite in the nanoflower, in agreement with the experimental extinction spectra in Fig.\ref{fig1}(b). \\

In order to relate the simulation information with the experiments performed at 1064 and 808 nm discussed below, we evaluated the average absorption cross-section, $\left \langle \sigma_{abs} \right \rangle$, from the simulated absorption efficiency values obtained for the different \textit{i} structures, $q_{abs}^i$, as $\left \langle \sigma_{abs} \right \rangle=\left \langle q_{abs}^i\right \rangle \pi a_{eff}^2$, where $a_{eff}$ is the effective radius of scatterers ~\cite{bohren2008absorption}. This information is detailed in Table \ref{tab:1}. This approach accounts for the sample size and shape heterogeneity in an \textit{effective} way.

\begin{table}[t]
  \centering
\begin{tabular}{c|cc|cc}
\hline
\multirow{2}{*}{Model} & \multicolumn{2}{c|}{$\lambda$= 1064 nm}                                                                         & \multicolumn{2}{c}{$\lambda$= 808 nm}                                                                         \\ \cline{2-5} 
                       & \multicolumn{1}{l|}{q$_{abs}\cdot 10^2$} & \multicolumn{1}{l|}{ $\sigma_{abs}\cdot 10^{6}(\mu \rm m^2$)} & \multicolumn{1}{l|}{q$_{abs}\cdot 10^2$} & \multicolumn{1}{l}{$\sigma_{abs}\cdot 10^{6}(\mu \rm m^2$)} \\ \hline
T1                     & \multicolumn{1}{c|}{3.02}              & 9.22                                                             & \multicolumn{1}{c|}{1.66}              & 5.09                                                           \\
t1                     & \multicolumn{1}{c|}{2.74}              & 8.93                                                             & \multicolumn{1}{c|}{1.48}              & 4.82                                                            \\
T2                     & \multicolumn{1}{c|}{3.40}              & 11.1                                                             & \multicolumn{1}{c|}{1.93}              & 6.30                                                            \\
t2                     & \multicolumn{1}{c|}{3.14}              & 11.5                                                              & \multicolumn{1}{c|}{1.79}              & 6.58                                                            \\
T3                     & \multicolumn{1}{c|}{3.66}              & 12.2                                                             & \multicolumn{1}{c|}{2.16}              & 7.21                                                            \\
t3                     & \multicolumn{1}{c|}{3.38}              & 13.5                                                              & \multicolumn{1}{c|}{1.95}              & 7.73                                                            \\ \hline
\end{tabular}
  \caption{Calculated values of absorption cross-section and absorption efficiencies of the selected particle models at the wavelengths of interest.}
  \label{tab:1}
\end{table}

\subsection{Nanothermometry of trapped MCNPs}

\subsubsection{Corner frequency-based nanothermometry}

\begin{figure*}[ht]\centering
\hspace*{-0.5cm}
\includegraphics[width=18 cm, angle= 0]{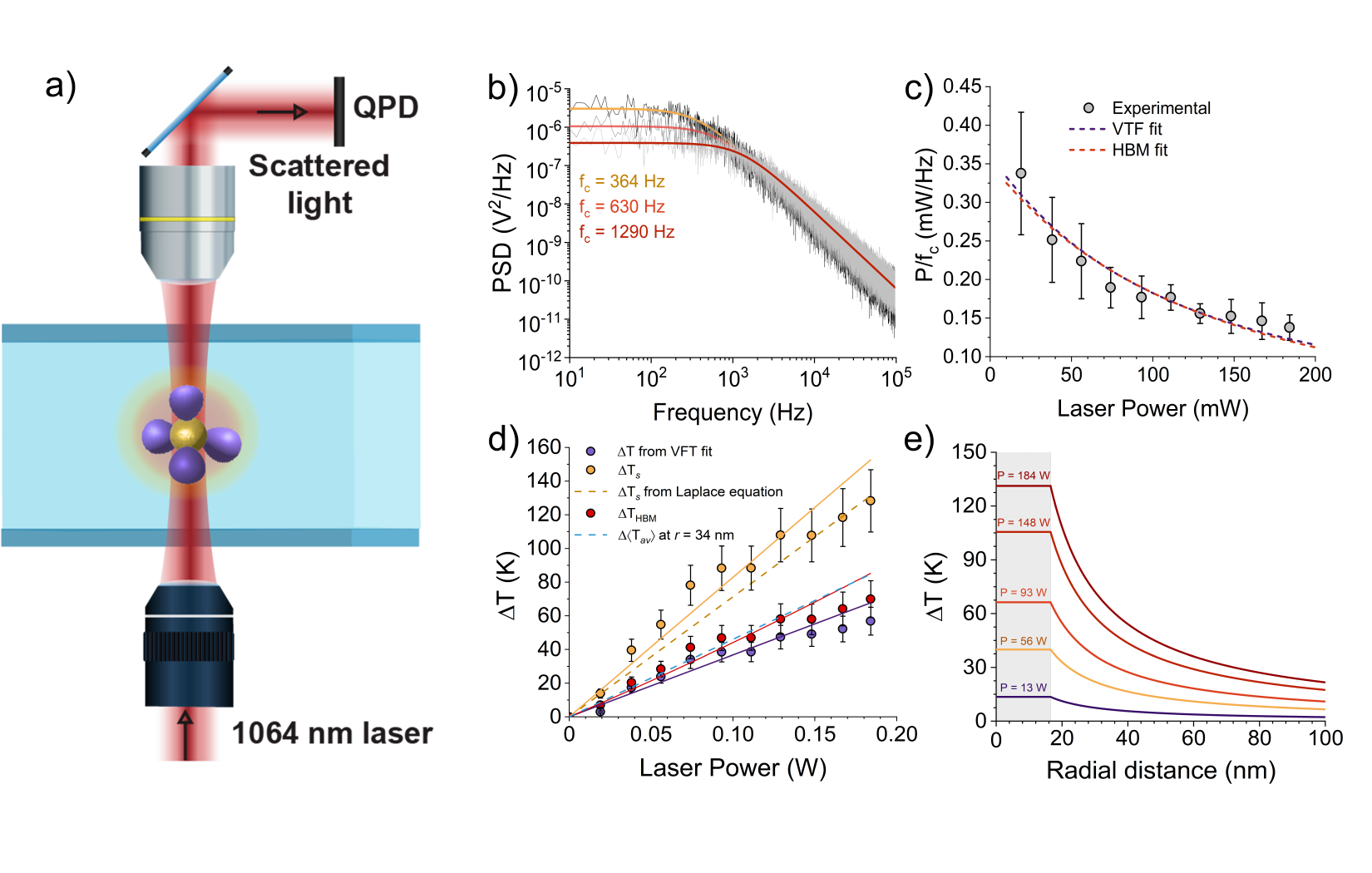}
\vspace*{-1.5cm} \caption{(a) Schematic representation of the inverted experimental setup for corner frequency-based nanothermometry. See sections Methods and B of Results and discussion for details. (b) Variation of the PSD with $P$. The black, dark grey and light grey curves stand for the PSD of a trapped MCNP at laser powers of 74, 111 and 167 mW, respectively. The Lorentzian fits required to evaluate $f_c$ are shown for exemplifying the experimental procedure to estimate $\Delta T$ upon laser heating. See the main text for details. (c) Experimental values of the laser power-to-corner frequency $ratio$ versus laser power. Error bars stand for the propagated error calculated from the standard deviation values evaluated over ten replicate measurements. The dashed lines denote the fit of the experimental data to Equation \ref{eq:Pf} (violet line) and to the HBM theory (red line), respectively. (d) Temperature increments as a function of the laser power. Symbols stand for the experimental data obtained from Equation \ref{eq:Pf} (violet circles), the surface temperature from HBM theory (red circles) and the \textit{effective} HBM theory (yellow circles), respectively. The continuous yelow and blue lines denote the results from the fittings, textit{i.e.}, $BP$ for $\Delta T$ as obtained from the VFT expression for $\eta$ (blue) and $B_{s}P$ for $\Delta T_{s}$ (yellow). The red continuous line results from the analytical value of $\Delta T_{HBM}$ evaluated from $T_0$ and $T_s$ following the prescriptions in Ref.\cite{rings10}. Results of $\Delta\left \langle T_{av}\right \rangle$ and $\Delta T_s$ calculated with the Laplace equation are represented as dashed lines following the color code of the full symbols. See the main text for details. (e) Temperature profiles obtained from the \textit{steady-state} Laplace equation at different values of the laser power. The absorbed power was evaluated under the strong confinement approximation according to Equation \ref{eq:power}. 
\label{fig-1}}
\end{figure*}

The first set of optical trapping and nanothermometry experiments were performed with a commercial optical tweezers setup described in the Experimental section and schematically represented in Figure \ref{fig-1}(a). Briefly, a single particle is trapped close to the beam waist of a tightly focused laser beam ($\lambda=1064$ nm). The forward scattered light carries information about the Brownian fluctuations of the particle around the trapping position, and it is collected and guided to a quadrant photodetector (QPD) that provides a voltage signal proportional to such fluctuations. Typically, that signal is analysed in terms of the power spectral density (PSD), \textit{i.e.}, the Fourier transform of its autocorrelation function \cite{gieseler2021optical}. For an overdamped system where a particle is in equilibrium with a thermal bath at absolute temperature $T$ and trapped in a parabolic potential well, the PSD depends on the frequency, $f$, following a Lorentzian function:

\begin{equation}\label{PSD}
    \text{PSD}(f)=\frac{D}{2\pi^2(f^2+f_c^2)}
\end{equation}
characterized by the diffusion coefficient of the particle $D=k_BT/\gamma$ and the corner frequency of the particle in the trap $f_c=\kappa/\gamma$, where $k_B$ is the Boltzmann constant, and $\gamma=6\pi \eta R_H$ is the Stokes friction coefficient for a sphere with hydrodynamic radius $R_H$ immersed in a fluid of viscosity $\eta$. Finally, $f_c$ is proportional to the stiffness of the trap, $\kappa$, which in turn is proportional to the power of the trapping laser, $P$. Therefore, one typically finds $f_c\propto P$ in the case of non-absorbing particles.\\

In our case, since the temperature of the liquid is unknown due to light absorption of the NPs, the standard calibration of displacement based on PSD cannot be used \cite{gieseler2021optical}. However, we can evaluate the effects of heating \textit{via} $f_c$, which can be determined without calibration since its measurement only depends on the internal clock of the electronic system. Therefore, 
the PSD of the Brownian fluctuations is fitted to a Lorentzian function from which $f_c$ can be evaluated (see Figure \ref{fig-1}(b)). The laser power-to-cutoff frequency $ratio$, $P/f_c$, should be proportional to the solvent viscosity, $\eta(T)$, that depends on the temperature $T$ in a nontrivial but well-known way. Accepting that the temperature can be linearly related to the laser power in the trap $P$ as $T=T_0+BP$ \cite{Paloma2018,gel}, where $T_0$ is the chamber temperature before laser heating, and using a Vogel–Fulcher–Tammann (VFT) dependence of the water viscosity on temperature from Ref.\cite{ddb}, the experimental data are non-linearly fitted to the function:

\begin{equation}
\label{eq:Pf}
    \frac{P}{f_c}=C\eta(T_0+BP),
\end{equation}

where $B$ and $C$ are fitting parameters. After this procedure, experimental results are shown in Fig. \ref{fig-1}(d) as violet symbols, together with the fitting curve (R$^2$=0.92, $\chi^2_\nu$=0.34). The fitted heating coefficient, $B$, was 368 $\pm$ 51 K$\cdot$W$^{-1}$, which is of the same order of magnitude as those obtained for different gold nanostructures in the literature \cite{Seol06,Andres-Arroyo15}. From this value, a temperature increment for the liquid around the particle can be obtained which can be as high as $\Delta T\simeq50$~K, as shown in Fig. \ref{fig-1}(d). However, the meaning of this temperature is not clear, since the trapped particle is hotter than the surrounding liquid, and a gradient of temperature is expected to develop \cite{rings10}, which can even be complicated by anomalies in the structure of water that may appear in this temperature range \cite{Lu20}. The temperature gradient, together with a concomitant viscosity one, affects the Brownian dynamics in the trap, whose evaluation can help us interpret the obtained results in a more reliable way. Hence,a system composed of a particle that is hotter than the environment due to a continuous energy input reaches a non-equilibrium steady state where the surface temperature of the particle $T_s$ gets a constant value that is higher than the surrounding liquid, and a temperature gradient close to it relaxes this value towards that of the thermal bath far from it. The theory of HBM shows that the dynamics of a particle in such a situation can be described by an \emph{effective} diffusion coefficient $D_{\rm HBM}=k_BT_{\rm HBM}/6\pi\eta_{\rm HBM}R_H$ given by an effective temperature $T_{\rm HBM}$ and an effective viscosity $\eta_{\rm HBM}$ that would return the measured diffusion coefficient of the very same particle placed in an equilibrium bath with those properties. Therefore, we apply the HBM theory to account for the out-of-equilibrium nature of the system by exploiting Equation \ref{eq:Pf} but employing the expression for $\eta_{\rm HBM}$ from Ref.\cite{rings10} instead (Eq. 8 in that reference). Since $\eta_{\rm HBM}$ incorporates the driving force responsible for the heat flux, the solvent temperature at the hydrodynamic boundary or surface temperature, $T_s$, at each laser power is obtained by fitting the data in Figure \ref{fig-1}(b). In a first approximation, the relation $T_s=T_0+B_sP$ should also hold, so this way we obtain the equivalent heating coefficient at the surface, $B_s= 830 \pm$ 110 K$\cdot$W$^{-1}$ according to the fitting (R$^2$=0.92, $\chi^2_\nu$=0.41).  Notably, values of $T_s$ above 100 $^{\circ}$C are observed at moderate and high laser powers without any manifestation of cavitation during the experiments. This effect has been previously reported and explained from different perspectives \cite{Paloma2018,superheat,heatgold}, from which the more plausible one under our experimental conditions is the formation of a surface tension-induced metastable and locally stretched fluid that prevents the formation of nano/microbubbles \cite{Paloma2018}. \\

In a second step, the \textit{effective} HBM temperature can be obtained following Ref.\cite{rings10}, using $\Delta T_{\rm HBM}\simeq \Delta T_s/2-[1-\rm ln$$(\eta_0/\eta_{\infty})]\Delta T_s^2/(24T_0)$, where $\eta_0$ is the viscosity in the absence of heating, \textit{i.e.}, at room temperature, and $\eta_{\infty}=0.0242631~\rm mPa\cdot s$. The new data are systematically higher ($\sim$25 $\%$ at the higher laser power value) than the \textit{Brownian} temperature increments derived above.  In other words, the effective temperature governing the dynamics of the trapped particle is higher than what is expected from the presumption that the dynamics follow classical Brownian motion, this effect being more apparent as the laser power increases. This happens because of the viscosity and temperature gradients within the nanoparticle boundaries and its surroundings. This observation may, at least partly, explain the discrepancies observed for different thermometric methods in Ref.\cite{Paloma2018}, where the effects derived from the HBM were ignored.
The confirmation that under strong heating conditions assuming a constant temperature provokes the underestimation of the center-of-mass motion of the particle sets traditional thermometric estimations and interpretations aside when manipulating efficient light-to-heat conversion nanomaterials.\\ 

A quick look at the so-obtained values of the temperature increments with \textit{P} allows one to observe that the expected linear trend is not fulfilled above intermediate laser powers. In principle, this outcome may suggest non-Fourier \textit{steady-state} behavior at energy fluxes in the trap of the order of 10$^{12}$ W$\cdot$m$^{-2}$. Monte Carlo simulations of heating nanoparticles have reported positive deviations of the surface temperature from the predictions of the heat transport constitutive equation, which are ascribed to non-continuous thermal conductivity and a mismatch between the solvent and the solid nanoparticle surface \cite{pnas}. Other similar effects have been experimentally observed with trapped nanoparticles, reporting a higher-than-expected variance of the instantaneous velocity of trapped Brownian particles due to a phase change in the water structure~\cite{Lu20}. Negative departure from the linearity has been also previously observed in some experimental setups \cite{Sahoo14}. However, we are not aware of theoretical treatments for describing these phenomena under optical trapping conditions. 
Furthermore, temperature-dependent hydration/hydration of the polymeric PEG coating \cite{PEG} might affect the local mass transport and heat exchange between the solvent layers, and water anomalies also takes place within the accessed temperature range \cite{water2}. Whether the captured experimental anomaly are due to either any of these contemplated or unreported effects or to a numerical artifact provoked by the strong variation of the viscosity with the temperature, that is more dramatically reflected in the high energy regime, is on the table now.\\

Additional insights into the experimental data can be provided by manipulating the analytical temperature profiles that were assessed as follows. The average absorption cross section determined above \textit{via} the discrete dipole scattering calculations are employed to evaluate the absorbed power, which is estimated under the proviso of strong confinement as \cite{fotonica}:

\begin{equation}
\label{eq:power}
P_{abs}=\frac{\left \langle \sigma_{abs} \right \rangle}{\frac{1}{2}\pi W_{0}^2}P,
\end{equation}
where $W_0=\lambda/(\pi \rm NA)$ is the waist radius and NA is the numerical aperture of the objective. The computed absorbed power was set as a seed for evaluating the temperature profiles, $T(r)$, outside the nanoparticle surface, $s$, from the analytical solution of the Fourier equation around an \textit{effective} spherical heating source at rest in a water bath, \textit{i.e.}, the \textit{steady-state} Laplace equation:


\begin{equation}
\label{eq: teo}
\Delta T(r)=\frac{P_{abs}}{4\pi \kappa_w r}\:\:\: \textup{for $r\geq s$},
\end{equation}
where $\kappa_w$=0.6 W$\cdot$K$^{-1}$$\cdot$m$^{-1}$ is the thermal conductivity of water and \textit{r} the radial distance from the particle center. Some examples of these profiles are shown in Figure \ref{fig-1}(d). Since the experimental value of \textit{B} implicitly considers the solvent heating, for comparative purposes this contribution was added \textit{ad hoc} according to the rate $B_w \sim $12 K$\cdot$W$^{-1}$ obtained in Ref.\cite{Peterman03} for water at 1064 nm. The average temperature increment at a distance $R=r-s$, $\Delta\left \langle  T_{av}(r)\right \rangle$, is then analytically obtained from the weighted volume integral:

\begin{equation}
\label{eq: average}
\Delta\left \langle  T_{av}(r)\right \rangle=\frac{\int_{s}^{r-s}4\pi r^2 \Delta T(r)dr}{\frac{4}{3}\pi (r^3-s^3)}+B_{w}P.
\end{equation}

Theoretical results obtained from Equation \ref{eq: average} are in excellent agreement with the experimental data when $\Delta\left \langle  T_{av}\right \rangle$ were obtained at a distance \textit{r} of 34 nm, just 5 nm away of the HD, as shown in Figure \ref{fig-1}(e). One may interpret the radial distance at which the \textit{steady-state} $\Delta\left \langle  T_{av}\right \rangle$ evaluated from the Laplace equation is equivalent to $\Delta T_{\rm HBM}$ as the point where the average kinetic energy of the molecules in the solvent bath stops being influenced by the HBM. This distance roughly coincides with the thickness of $\sim$70 monolayers of water, from which one may deduce the solvent volume required to subdue the strong effect of temperature gradients by thermalization. Moreover, assuming the generated heat power needed to increase the bath temperature to the effective $T_{\rm HBM}$ is wholly absorbed within this solvent volume, we estimated the time required to reach the HBM effective temperature is $\sim$200 ns. 

\subsubsection{Emission-based nanothermometry}

\begin{figure*}[ht!]
\hspace*{-0.5cm}
\includegraphics[width=18 cm, angle= 0]{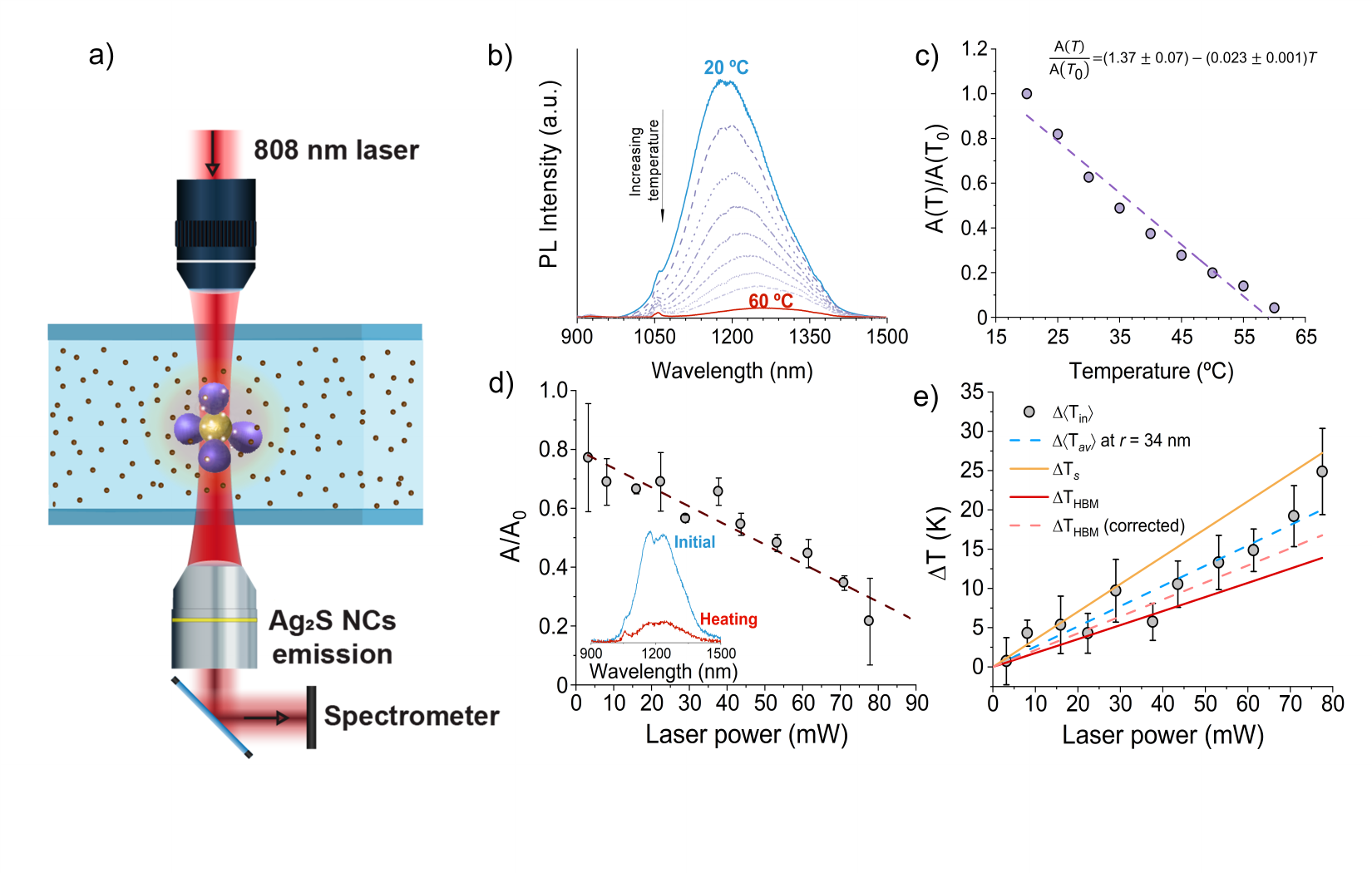}
\vspace*{-1.5cm}  \caption{(a) Scheme of the trapping setup for emission-based nanothermometry. (b) Variation of the photoluminescence spectra of a Ag$_2$S NCs suspension with the temperature. (c) Calibration curve of the normalized photoluminescence spectra area of the Ag$_2$S NCs suspension \textit{versus} the externally controlled temperature. Error bars are smaller than the symbol size. (d) Normalized photoluminescence spectra area as a function of the laser power (808 nm) as measured in the trapping experiments. The inset shows an example of the raw detected spectra at the lowest (initial) and highest (final) laser powers. Error bars are calculated from the standard deviation of, at least, triplicated experiments. (e) Experimental $\Delta\left \langle  T_{in}\right \rangle$ temperature obtained from the combination of the experimental data in bulk Ag$_2$S NCs suspensions in (c) with the linear calibration in (d). Error bars are propagated from those of (d) and the determination error of the calibration in (d). Theoretical results obtained for $\Delta\left \langle  T_{av}\right \rangle$ to a radial distance of 21.5 nm from the NP surface is also included as a dashed line. The water and nanothermometers heating rate at 808 nm was set at $B_{nt}$ = 42 K$\cdot$W$^{-1}$ according to the blank experiments detailed in the Methods section.
\label{fig-3}}
\end{figure*}

The second set of experiments was accomplished by monitoring the NIR emission of a dispersion of nanothermometers in which the NFs are embedded. The schematic representation of the trapping and photoluminescence (PL) detection setup is sketched in Fig. \ref{fig-3}(a), and detailed in the Experimental Section. In previously published results, it was shown that the intensity of the NIR emission at $\sim$1200 nm of Ag$_2$S nanocrystals (NCs) can act as reversible nanothermometers \cite{YingliAg2S2020,YingliAg2S2022}, and are here used for the first time for nanothermometry purposes in optical trapping. Briefly, we employed a commercial suspension of Ag$_2$S-PEG-COOH nanothermometers that, prior to the trapping experiments, were calibrated following a procedure in which the temperature dependence of the emission of an NCs dispersion is taken as the transduction signal. In Figure \ref{fig-3}(b) this variation is presented. The changes in the area of the emission band with temperature were used to determine the thermal sensitivity of the Ag$_2$S NCs that followed the linear relation A($T$)/A($T_0$)=1.37$_7$-0.023$_1$$T$ (R$^2$ = 0.96) as shown in Fig. \ref{fig-3}(c), where A($T$)/A($T_0$) is the $ratio$ of the spectral area at temperature $T$, A($T$), and that at a reference temperature $T_0$, A($T_0$). From this calibration, it can be concluded that the use of Ag$_2$S NCs for nanothermometry is restricted to temperatures up to $\sim$60 $^o$C. Then, an aqueous colloidal suspension of NPs and Ag$_2$S NCs was placed in a homemade single-beam optical trapping setup, described in the Experimental Section, that uses an 808 nm linearly polarized single mode fiber--coupled laser diode. The laser was used for both trapping the nanoflowers and exciting the NCs in their closer surroundings. The NCs emission was monitored with a NIR spectrometer fiber--coupled to the system. The normalized photoluminescence area of the NCs dispersion was measured as a function of the laser power (808 nm) in the trapping setup as shown in Fig. \ref{fig-3}(d). The inset shows an example of the raw detected spectra at the lowest and highest laser powers. Finally, the spectral data were converted to temperature data using the calibration line obtained for the nanothermometers suspension. The final results are collected in Fig. \ref{fig-3}(e) as full symbols. \\

It must be stressed that our measurement through the emission changes in the NIR-spectra upon heating reflects the average temperature of the Ag$_2$S nanocrystals contained in a liquid volume around the trapped particle, $\left \langle  T_{in}\right \rangle$, and this is not a direct observable evincing the trapped nanoparticle dynamics. Particularly, the experimental values of $\Delta\left \langle  T_{in}\right \rangle$ reflect real differences in the average temperature around the heated nanoparticle. As shown as a dashed blue line in Figure \ref{fig-3}(e), we found that the values of $\Delta\left \langle  T_{av}(r)\right \rangle$ obtained from the Laplace equation at $r$ = 34 nm, including the heating contribution of the solvent and the nanothermometers, $B_{nt}$ (see Methods), are in excellent agreement with the $\Delta\left \langle  T_{in}\right \rangle$ data. Therefore, the question that arises after these results is whether the radial distance required to obtain an average temperature compatible with the spectroscopic data is representative of the effective solvent boundary affected by the HBM. Although the overall agreement between the experimental $\Delta\left \langle  T_{in}\right \rangle$ data and the theoretical values of $\Delta\left \langle  T_{av}(r)\right \rangle$ at a radial distance is consistent with the experiments at 1064 nm, solving this fundamental issue is cumbersome from any angle. In the next section, we try to address this issue directly by comparing both data sets.

\subsubsection{Comparison of the two data sets and discussion}

In this section, we try to find a connection between the two types of performed measurements. Some hints can be provided by pondering that, in a first approximation, both experiments should be fully equivalent in terms of heat generation if the absorbed power is considered a common variable, regardless of the trapping wavelength. Provided the nice agreement found in the experiments performed at 1064 nm, we evaluated the surface temperature from the Laplace equation employing the $\left \langle \sigma_{abs} \right \rangle$ at 808 nm obtained $via$ DDSCAT calculations, from which $T_{HBM}$ can be derived. The analytical results of both temperatures are plotted as continuous (yellow for $\Delta T_s$ and red for $\Delta T_{\rm HBM}$) lines in Figure \ref{fig-3}(e). The agreement between $\Delta\left \langle  T_{av}(r)\right \rangle$ and $\Delta T_{\rm HBM}$ is highly remarkable, singularly considering that no contributions of the solvent and/or nanothermometers are contemplated in the HBM temperature. Under the naïve assumption that this contribution is additive to the theoretical value of $\Delta T_{\rm HBM}$, we obtained the dashed orange line in Figure \ref{fig-3}(e) (denoted as $\Delta T_{\rm HBM}$ corrected), which is in even closer agreement with $\Delta\left \langle  T_{av}(r)\right \rangle$. This is a manifestation that $\Delta\left \langle  T_{av}(r)\right \rangle$ is closely related with $\Delta T_{\rm HBM}$ as evaluated from the trapped particle heating once the contribution of its surroundings is accounted for by, in a first approximation, simple addition.\\

\begin{figure}[ht!]
\hspace*{-0.5cm}
\includegraphics[width=8.5 cm, angle= 0]{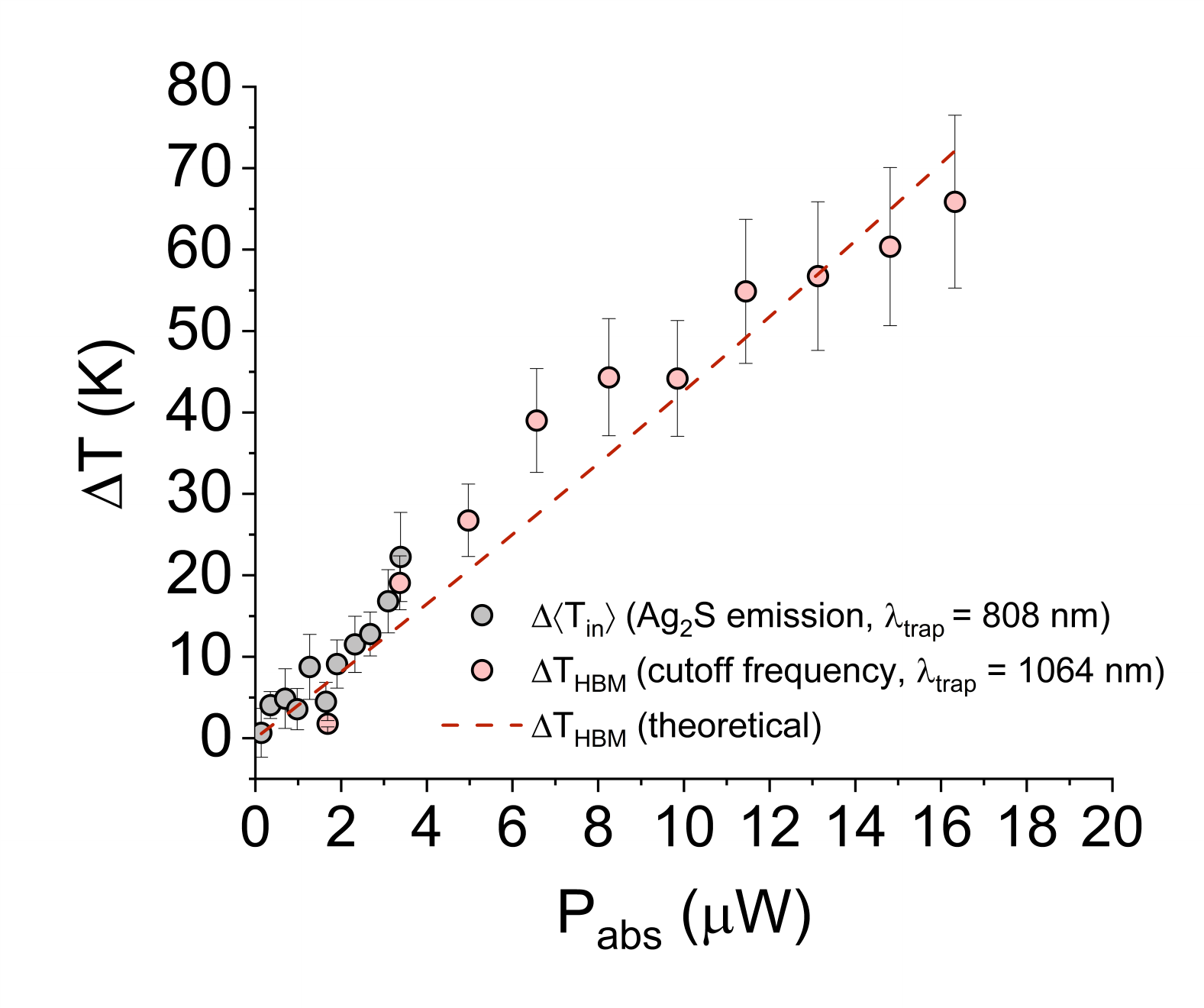}
\vspace*{-0.5cm}  \caption{Comparison between $\Delta T_{\rm HBM}$ increments obtained with the cutoff frequency (yellow symbols) and $\Delta\left \langle  T_{in}\right \rangle$ as obtained with the emission of Ag$_2$S NCs (grey symbols). The continuous red line stands for the theoretical calculation of $\Delta T_{\rm HBM}$ derived from the HBM theory assuming a temperature profile as defined in Equation \ref{eq: teo}.
\label{fig-4}}
\end{figure}

Following this reasoning, the plot of the values of $\Delta T$ coming from both sets of experiments against $P_{abs}$ should overlap if the extracted $\Delta T$ values have some physical resemblance between them. In other words, we can disclose if $\Delta\left \langle  T_{in}\right \rangle$ is comparable with $\Delta T_{\rm HBM}$ in the whole power regime and, consequently, whether remote-non-trapped thermometers are reliable tools to described the HBM dynamics within their thermal working range. This comparison is presented in Fig. \ref{fig-4}, where the theoretical value of $\Delta_{\rm HBM}$ is also included. Notice that one can evaluate the theoretical $\Delta T_{s}$ value from the Laplace equation independently of the concrete initial temperature $T_{0}$ but, since $T_{\rm HBM}$ is a function of $T_0$ and our experimental values are different in $\sim$2 K in each setup, the continuous line representing the theoretical $\Delta T_{\rm HBM}$ value has been calculated for the average $T_{0}$. The agreement between experimental and theoretical data is such that the RMSE is $\sim$6~K. Besides, the contribution of the nanothermometers has been subtracted from the experiments at 808 nm for the sake of comparison. The representation of those sets of data in the absorbed power range of 0--20 $\mu$W indicates that the values of $\Delta\left \langle  T_{in}\right \rangle$ derived from auxiliary aide-emissive nanothermometers nicely correlate, within experimental uncertainties, with those defining the particle dynamics in the framework of the HBM theory when the laser power was appropriately scaled with the absorption cross-section of the nanoparticles. This result indicates both negligible effects of the nanothermometers on the physical properties of the heat source and, $viceversa$, and that the Fourier constitutive equation remains applicable in the low fluence regime at the nanoscale, even for estimating the HBM temperature.

\section{conclusions}
To sum up, the temperature and heat generation by absorbing anisotropic hybrid nanostructures under optical trapping conditions were experimentally and theoretically achieved on the basis of:\\

(i) \emph{The evaluation of the balance of the optical forces on the confining trap as a function of the laser power}. The experimental temperature data were extracted from the expected dependence of the power-to-corner frequency \textit{ratio} with the \textit{effective} medium viscosity as established in the HBM theory. The evaluated heating parameter in the classical Brownian approximation is in correspondence with previously reported data. In contrast, the HBM method led to temperature increments systematically higher than those obtained from the classical Brownian motion vision of a particle dynamics immersed in a thermal bath. This fact points to the necessity of applying the sparingly employed HBM theory to extract valid information from thermometric measurements when handling highly absorbing particles and question the traditional meaning of \textit{temperature} as far as single particle nanothermometry is concerned. It was also verified that at low energy fluxes the Fourier equation remains applicable to obtain the HBM \textit{temperature} from the directly estimated value of \textit{T$_s$}.\\

(ii) \emph{The calibration of a colloidal suspension of NIR emissive nanothermometers embedding the trapped nanoparticles}. Average internal temperature increments within the trapping volume have been accessed directly by the near-infrared emission of Ag$_2$S NCs acting as NIR nanothermometers. The values obtained within this approach constitute a sort of average temperature and it was found that it might be resembling the HBM temperature after correcting the heating contribution of the environment. This asseveration was demonstrated by proving that by merging the experimental data coming from both sets of results for the dependence of $\Delta T$ with $P$, remarkably obtained in two different laboratories and with two different trapping setups operating at different trapping wavelengths, they collapsed when plotted against the absorbed power.\\

These findings represent a robust standpoint of the temperature measurements in single-particle systems in relation to the experimental route from which the heating properties are derived. That said, while experimental pieces of evidence indicate that the internal temperature determined from the appropriate calibration of NIR-emissive nanothermometers capture that dictating the particle dynamics (\textit{i.e.}, the HBM temperature), no rigorous theoretical validation of such conclusion has been reported to date. Further applications of the protocol initiated in this work might yield a reinterpretation or validation of the trapping experiments on self-emissive nanothermometers \cite{nanot}. On a final note, in light of the importance of the HBM theory to elucidate nanothermometric data coming from optical trapping setups, it seems urgent to generalize it to incorporate the local heating contribution of both the nearby particles and solvent and the effects of the particle softness \cite{gel} into the formalism.

\section*{Methods}

\textbf{Chemical reagents.}
Gold (III) acetate, iron (III) chloride, sodium oleate, oleic acid 99\%, oleylamine, 1,2-hexadecanediol, gallic acid, polyethyleneglycol (PEG) 3000 Da, 1-octadecene, triethylamine, 4-dimethylaminopyridine (DMAP), and ethanol 99\% were obtained from Sigma Aldrich. Dimethyl sulfoxide (DMSO), toluene, acetone, hexane, chloroform, dichloromethane (DCM), and tetrahydrofuran (THF) were supplied by Acros organics. All reagents were used as received without further purification. Milli-Q water (18.2 M$\Omega\cdot$cm) were obtained from a Millipore system. Ag$_2$S NCs were obtained from NIR Optics Technology (average size, $\sim$9 nm).\\

\textbf{Seed-growth synthesis of Au@Fe$_3$O$_4$ nanoflowers.}\\

\textit{Synthesis of iron (III) oleate}. The synthesis was done following Ref. \cite{Pernia17}. Briefly, a mixture of 10.8 g of iron (III) chloride (40 mmol) and 36.5 g of sodium oleate (120 mmol) were dissolved in a solvent mixture with 80 mL of ethanol, 60 mL of distilled water and 140 mL of hexane. The resulting solution was heated to 60 ºC for 4 h in hexane reflux under inert atmosphere. The reaction was then cooled down to room temperature. The organic phase was washed 3 times with distilled water and the hexane was evaporated in the rotavapor.\\
 
\textit{Synthesis of the gold seeds}. The synthesis was done following Ref.\cite{Tancredi19} with some modifications. 50 mg of gold (III) acetate were dissolved in a mixture of 0.8 mL oleic acid, 0.6 mL of oleylamine, 100 mg of 1,2-hexadecanodiol and 5 mL of 1-octadecene. The mixture was heated under vacuum to 120 ºC with a heating rate of 5 ºC$\cdot$min$^{-1}$, and kept at this temperature for 30 min. After cooling down, it was washed twice with 36 mL of a mixture of ethanol/acetone (1:1, v:v) and centrifuged at 8000 rpm for 20 minutes. Then, the gold NPs were dispersed in 10 mL of hexane.\\

\textit{Synthesis of Au-Fe$_3$O$_4$ nanoflowers}. The synthesis was conducted as previously described by some of us \cite{Christou22}. 1 mL of Au NPs was mixed with 0.63 mL of oleylamine, 0.66 mL of oleic acid, 0.645 mL of 1,2-hexadecanodiol, 10 mL of 1-octadecene and 0.125 g of iron (III) oleate. The mixture was heated under vacuum to 120 ºC and kept at this condition for 20 min. Then, the temperature was raised to 200 ºC and kept at this temperature for 120 min. The temperature was raised again to 315 ºC with a heating rate of 5 ºC$\cdot$min$^{-1}$ and kept at this temperature for 30 min (growing of iron). After cooling it down, it was washed with a mixture of ethanol/acetone (1:1) and centrifuged . This step was done twice. Then, the gold NPs were dispersed in 10 mL of hexane.\\

\textbf{Functionalization of the Au-Fe nanoflowers.}\\

The synthesis of the appropriate PEGylated ligand and the subsequent ligand exchange process were conducted as described previously by some of us in Ref.\cite{Pozo20}. In the following we provide with the more important information.\\

\textit{Synthesis of the PEGylated ligand}. First, we proceed with a dropwise addition of DCC (1 g in 5 mL of THF) to a solution of 3 g of PEG, 170 mg of gallic acid and 24 mg of DMAP in 100 mL of THF and 10 mL of DCM. The resulting mixture was stirred overnight at room temperature and, finally, it was filtered through a filter paper and the solvents were rota-evaporated. \\

\textit{Ligand Exchange.} The ligand exchange procedure was perfomed following \cite{Pozo20}. In short, in a glass vial a solution containing 1 mL of NPs (10 g/L of Fe), 1 mL of the gallol-PEG$_n$-OH in a concentration of 0.1 M in CHCl$_3$ and 50 $\mu$L of triethylamine was added. The mixture was ultrasonicated for 1 h and kept 4 h at 50 ºC. At this point, it was diluted with 5 mL of toluene, 5 mL of Milli-Q water and 10 mL of acetone. Then, it was shaked and the nanoparticles were transferred to the aqueous phase. After that, the aqueous phase was collected in a round-bottom flask and the residual organic solvents were rota-evaporated. Then, the gallol derived NPs were purified in centrifuge filters with a molecular weight cutoff of 100 kDa at 450 rcf. In each centrifugation, the functionalized NPs were re-suspended with Milli-Q water. The purification step was repeated several times until the filtered solution was clear. After the purification, the gallol derived NPs were re-suspended in PBS buffer. Finally, to improve the monodispersity and remove aggregates, this solution was centrifuged at 150 rcf for 5 min and it was placed onto a permanent magnet (0.6 T) for 5 min.\\

\textbf{Physicochemical characterization of the Au-Fe nanoflowers.} The size distribution were obtained by TEM imaging carbon-coated copper grid in which a sample suspension of $\sim$1 g/L of (Fe+Au) are dropwise deposited. The images were acquired on a FEI Tecnai G2 Twin microscope working at 100 kV. Size histograms were then calculated by averaging the characteristic dimensions of 100 nanoparticles with the aid of the ImageJ free software. The extinction spectra were recorded in a Jenway Series 67 spectrophotometer.. The measurement were performed in a very dilute suspension to avoid potential collective effects affecting the scattering contribution due to the potential aggregation of the nanoparticles.\\

\textbf{Piriform curve for particle modelling.} The geometry used to implement the simulations is based on the revolution of the piriform curve, which only depends on the value of a parameter $\beta$, and is given in parametric form by the following expression:
\begin{multline}
\label{eq: piriform}
\beta^2=(x^2+y^2)(1+2x+5x^2+6x^3+6x^4\\+4x^5+x^6-3y^2-2xy^2+8x^2 y^2\\+8x^3 y^2+3x^4 y^2+2y^4+4xy^4+3x^2 y^4+y^6).
\end{multline}

\textbf{Optical trapping at 1064 nm.} The experiments at 1064 nm were performed in a NanoTracker-II optical tweezers (JPK-Bruker) device. In this setup, the infrared laser was tightly focused by a high numerical aperture objective (63$\times$, NA$_b$=1.2) to a diffraction-limited spot where trapping occurs. The forward-scattered light is collected by a second objective and guided to a QPD to record the X-Y traces that generate a voltage signal sampled at a frequency of 50 kHz. This voltage is proportional to the displacement of the particle inside the optical trap. Spurious bright flashes in the video were assumed to be a consequence of a multiple trapping event and, in such cases, the PSD and the corresponding $f_c$ were discarded.\\

\textbf{Optical trapping at 808 nm.} The optical trapping experiments were performed in a homemade single-beam optical trapping setup. A drop of aqueous suspensions of MCNPs and NCs, previously stirred to avoid clusters, was pipetted into a 120 $\mu$m height and 13 mm diameter micro-chamber that was placed in the optical tweezers setup. A linearly polarized 808 nm single-mode fiber--coupled laser diode was focused into the chamber containing the sample by using a LCPLN 100$\times$ IR Olympus microscope objective with a numerical aperture of 0.85 that leads to a spot size of $\sim$0.63 $\mu$m. The tightly focused laser beam was used for both trapping the nanoflowers and exciting the NCs in its closer surroundings. Real time optical imaging of the NCs was achieved by coupling a white LED, focused on the sample by a 40$\times$ objective lens, and using a CMOS camera incorporated into the system. The lower objective lens is used as a light condenser, but also serves as a collector lens to focus the NCs emission into a OceanInsight NIR spectrometer, fiber--coupled to the system.\\

\textbf{External and internal Ag$_2$S NCs calibration.} For the external calibration of the Ag$_2$S in bulk dispersions, the temperature of the sample was controlled by a Linkam PE120 stage ($\pm$ 0.1 ºC). The NCs dispersion was excited with a fiber-coupled 808 nm single-mode diode laser while the NIR emission spectra were collected with an OceanInsight high-performance NIR Spectrometer (900--1700 nm). A high quality long-pass filter (750 nm) was placed in the detection path so that the noise level registered by the camera did not exceed 0.5 $\%$ of the signal generated by the Ag$_2$S NCs. While the heating contribution of water in the trap might be set to  $B_{w}$ = 3 K$\cdot$W$^{-1}$ \cite{water}, the contribution of the Ag$_2$S NCs is unknown. To evaluate its contribution, a set of experiments on the colloidal suspension under trapping conditions (but without trapped MCNPs) were performed. The temperature increments were evaluated from the internal calibration curve, and the solvent plus NCs heating coefficient, $B_{nt}$, was determined from the slope of the $\Delta T$ against laser power measurements. The calibration curve is shown in Figure S4. The outcome of these experiments is $B_{nt}$ $\approx$ 42 K$\cdot$W$^{-1}$.  \\

 \par 

\medskip
\textbf{Acknowledgements} \par 
The authors acknowledge funding from the Ministerio de Ciencia e Innovación of Spain (grants PID2022-136919NA-C33, PID2021-127427NB-I00, PID2020-118448RBC21, PID2019-105195RA-I00, CNS2022-135495, and TED2021-129937B-I00), from the Ministerio de Economía, Industria y Competitividad of Spain (grant CTQ2017-86655-R) and from FEDER/Consejería de Transformación Económica, Industria, Conocimiento y Universidades of Andalucía (grants P18-FR-3583 and P20$\_$00727/PAIDI2020). H.C. and R.A.R. gratefully acknowledge funding from HORIZON-MSCA-2021-PF-01 Grant agreement ID: 101065163. E.O.R gratefully acknowledges the financial support provided by the Spanish Ministerio de Universidades, through the FPU program (FPU19/04803) and C. C. thanks to the Consejería de Salud y Familias (Junta de Andalucía) for his senior postdoctoral grant (RH-0040-2021).

\clearpage
\medskip
\textbf{Bibliography}
\bibliographystyle{ieeetr}



\end{document}